\documentclass[prd,twocolumn,superscriptaddress,nofootinbib]{revtex4}

\usepackage{amsfonts}
\usepackage{amsmath}
\usepackage{amssymb}
\usepackage{amsthm}
\usepackage{bm}
\usepackage{dcolumn}
\usepackage{epsfig}
\usepackage{graphicx}
\usepackage{graphics}
\usepackage[latin1]{inputenc}
\usepackage{latexsym}
\usepackage{rotating}
\usepackage{hyperref}

\newcommand\be{\begin{equation}}
\newcommand\ba{\begin{eqnarray}}
\newcommand\ee{\end{equation}}
\newcommand\ea{\end{eqnarray}}

\newcommand{\GR}{{\mbox{\tiny GR}}}

\begin{document}
\title{Binary Pulsar Constraints on the Parameterized post-Einsteinian Framework}

\author{Nicol\'as Yunes}
\affiliation{Department of Physics, Princeton University, Princeton, NJ 08544, USA.}

\author{Scott A.~Hughes}
\affiliation{Department of Physics and MIT Kavli Institute, Cambridge, MA 02139, USA.}

\date{\today}

\begin{abstract}

We constrain the parameterized post-Einsteinian framework with binary
pulsar observations of orbital period decay due to gravitational wave
emission. This framework proposes to enhance the amplitude and phase
of gravitational waveform templates through post-Einsteinian
parameters to search for generic deviations from General Relativity in
gravitational wave data. Such enhancements interpolate between General
Relativity and alternative theory predictions, but their magnitude
must be such as to satisfy all current experiments and
observations. The data that currently constrains the parameterized
post-Einsteinian framework the most is the orbital period decay of
binary pulsars. We use such observations to place upper limits on the
magnitude of post-Einsteinian parameters, which will be critical when
gravitational waves are detected and this framework is implemented.
\end{abstract}


\maketitle

{\emph{Introduction}}. Gravitational waves (GWs) will allow us to
learn about the gravitational interaction in regimes that are
currently inaccessible by more conventional, electromagnetic
means. Binary black hole and neutron star mergers, for example, lead
to gravitational fields that are intensely strong and highly
dynamical, a regime where General Relativity (GR) has not yet been
tested. GW theorists and data analysts will need to be able to make
quantitative statements about the confidence that a certain event is
not just a GW detection but one consistent with GR.

The parameterized post-Einsteinian (ppE) framework~\cite{Yunes:2009ke}
was devised precisely for this purpose: to search for statistically
significant GR deviations or {\emph{anomalies}} in GW data and, in
their absence, to quantify the degree of belief that a GW event is
purely described by GR. This framework enhances the waveform templates
used in matched-filtering through parameters that characterize GR
deformations. In practice, this is achieved by adding to GR templates
amplitude and phase correction, with magnitudes depending on certain
ppE parameters.

Any framework that modifies GR must comply with Solar System and
binary pulsar observations. These measurements already strongly
constrain GR deviations in weak and moderately strong fields. The ppE
framework was constructed on a maxim of compliance with current
observations, which can be enforced by requiring that the magnitude of
the ppE correction be such as to satisfy current constraints. Until
now, this maxim had not been quantitatively enforced because it was
thought that it would be difficult to relate the ppE deformations to
Solar System or binary pulsar observations.
 
We have here found a relatively simple way to relate the ppE framework
to current experiments. As shown in~\cite{Yunes:2009ke}, modifications
to the dissipative and conservative sectors of the theory lead
structurally to similar ppE corrections to the waveform. We find that
to constrain the ppE framework with current experiments at least
initially, it suffices to consider dissipative corrections only, while
keeping the conservative sector unmodified. Such dissipative
corrections modify the amount of orbital binding energy carried away
by GWs, which affects directly the orbital period decay in binary
pulsars.

The relatively recent discovery of the binary pulsar, PSR
J0737-3039~\cite{Lyne:2004cj}, has provided particularly powerful GR
tests~\cite{Kramer:2006nb}. This pulsar is highly relativistic, with
an orbital period of about 2 hours, and has an orbital geometry
favorable for measuring quantities such as the Shapiro delay with
sub-percent precision. Such data has been recently used to constrain
alternative theories of gravity to new levels~\cite{Yunes:2008ua}.

In this paper, we relate such sub-percent accurate measurements of the
orbital decay of PSR J0737-3039 to constrain the ppE framework and its
templates. Because of the structure of the ppE correction to GWs,
these constraints are {\emph{relational}}, i.e.,~they are of the form
$|\gamma| f^{c} \leq F(\delta,\vec{\lambda})$, where $(\gamma,c)$ are
ppE parameters, $f$ is the GW frequency and $F(\delta,\vec{\lambda})$
is some function of the accuracy $\delta$ to which the orbital decay
has been measured and system parameters $\vec{\lambda}$, such as the
mass ratio and total mass of the binary. Thus, given a value for $c$,
the magnitude of $\gamma$ is constrained by binary pulsar observations
to be less than some number related to $\delta$, $f$ and
$\vec{\lambda}$. The relational constraint found in this paper will be
crucial in the implementation of the ppE framework in a realistic data
analysis pipeline once GWs are detected. In the rest of this paper, we
follow mostly the conventions of~\cite{Misner:1964} with geometric
units $G=c=1$.

{\emph{Basics of the ppE Framework}}. The main GW observable is the
so-called response function, which describes how an interferometer
reacts to an impinging GW. In GR, this function is given by
\be
h_{\GR}(t) \equiv F_{+} h_{+}^{\GR}(t) + F_{\times} h_{\times}^{\GR}(t)\,,
\label{resp-func}
\ee
where $F_{+,\times}$ are beam-pattern functions and
$h_{+,\times}^{\GR}$ are the plus and cross GW polarizations, built
from contractions of the metric perturbation with certain polarization
tensors~\cite{Misner:1964}. For quasi-circular binaries, these
polarizations can be Fourier transformed in the stationary-phase
approximation~\cite{Flanagan:1997sx,Droz:1999qx,Yunes:2009yz} to yield
\ba
\label{hp-hc}
\tilde{h}_{+}^{\GR} &=& - \frac{{\cal{M}}}{D_{L}} \frac{u^{2/3}}{\sqrt{2 \dot{F}}} \left(1 + \cos^{2}\iota \right) e^{-i \left(\Psi_{\GR} + \pi/4 - 2 \beta \right)}\,,
\nonumber \\
\tilde{h}_{\times}^{\GR} &=& - \frac{{\cal{M}}}{D_{L}} \frac{u^{2/3}}{\sqrt{2 \dot{F}}} \left(2 \cos\iota \right) e^{-i \left(\Psi_{\GR} - \pi/4 - 2 \beta\right)}\,,
\ea
where $(\iota,\beta)$ are the inclination and polarization angles,
$D_{L}$ is the luminosity distance from source to observer and
$\dot{F}$ is the rate of change of the orbital frequency due to GW
emission. This frequency is defined as $F \equiv (1/2\pi) \dot{\Psi}$,
where $\Psi$ is the orbital phase, and it is also equal to half the
Fourier or GW frequency $f$, i.e.,~$F = f/2$. The quantity
$\Psi_{\GR}$ is the GR GW phase in the Fourier domain, which can be
computed via
\be
\Psi_{\GR}(f) = 2 \pi \int^{f/2} \frac{F'}{\dot{F}'} \left(2 - \frac{f}{F'} \right) dF'\,.
\label{GW-phase}
\ee
The quantity $u \equiv \pi {\cal{M}} f$ is a dimensionless frequency
parameter, where ${\cal{M}} = \eta^{3/5} m$ is the chirp mass, with
$\eta = m_{1} m_{2}/m^{2}$ the symmetric mass ratio and $m = m_{1} +
m_{2}$ the total mass. From Eq.~\eqref{resp-func}, it follows that the
Fourier transform of the response function in the stationary phase
approximation is simply $\tilde{h}_{\GR} = F_{+} \tilde{h}_{+}^{\GR} +
F_{\times} \tilde{h}_{\times}^{\GR}$.

The ppE framework proposes that one enhances the GR response function
via an amplitude and a phase correction. In the Fourier domain and in
the stationary phase approximation, one can parameterize the response
function for a GW from an unequal-mass, binary, quasi-circular
inspiral as~\cite{Yunes:2009ke,ppE-tbp}
\be
\tilde{h} = \tilde{h}_{\GR} \left(1 + \alpha \; \eta^{c} u^{a} \right) e^{i \beta \; \eta^{d} u^{b}}\,,
\label{ppE-h}
\ee
where $(\alpha, c,a)$ are ppE amplitude parameters and $(\beta,d,b)$
are ppE phase parameter. Such a correction arises generically if one
modifies $\dot{F} = \dot{E} \; (dE_{\rm b}/dF)^{-1}$, which in turn
can arise either due to a modification to the GW luminosity $\dot{E}$
(the dissipative sector) or to the orbital binding energy $E_{\rm b}$
(the conservative sector). As explained in~\cite{Yunes:2009ke}, this
degeneracy breaks the one-to-one mapping from a ppE waveform
modification to a specific alternative theory, as one cannot tell
whether the change arose in the dissipative or conservative sector.

{\emph{Gravitational Wave Luminosity}}. We now compute the energy
carried by ppE GWs. As is clear from Eq.~\eqref{hp-hc}, the GW
amplitude depends on $\dot{F}$, which by the chain rule can be related
to $\dot{E}$ as explained below Eq.~\eqref{ppE-h}. We can construct
$\dot{E}$ directly from $h_{+}$ or $h_{\times}$ via
\be
\dot{E} = \frac{\pi}{2} f^{2} D_{L}^{2} \dot{f}_{\rm GR} \int d\Omega 
\left(\left| \tilde{h}_{+} \right|^{2} + \left| \tilde{h}_{\times} \right|^{2}\right)\,,
\label{dotE-amp-ppE}
\ee
where $\dot{f}_{\GR}$ is the rate of change of the GW frequency, and
$d\Omega = \sin{\iota} \; d\iota \; d\beta$ integrates over the
$(\iota,\beta)$ dependence of the waveform. Notice that
Eq.~\eqref{dotE-amp-ppE} agrees with Eq.~$(2.38)$
in~\cite{Flanagan:1997sx}. Substituting for $\tilde{h}$ using
Eq.~\eqref{ppE-h}, we find
\be
\dot{E} = \dot{E}_{\GR} \left|1 + \alpha \; \eta^{c} u^{a} \right|^{2}\,,
\label{dotE-amp-ppE2}
\ee
where $\dot{E}_{\GR}$ is the GR expectation for the GW luminosity:
\be
\dot{E}_{\GR} = \frac{\pi}{2}  \dot{f}_{\GR} f^{2} D_{L}^{2} \int d\Omega 
\left( \left| \tilde{h}_{+}^{\GR} \right|^{2} + \left|  \tilde{h}_{\times}^{\GR} \right|^{2} \right)\,,
\ee

One can also obtain an expression for the GW luminosity in terms of
the GW phase only, as this also depends on $\dot{F}$ as shown in
Eq.~\eqref{GW-phase}. Noting that $d^{2} \Psi/df^{2} = \pi
\dot{F}^{-1}$, we can write the GW luminosity as
\be
\label{dotE-phase-ppE}
\dot{E} = - \frac{1}{6} \, \dot{f}_{\GR}^{2} \,  {\cal{M}}^{2} \,  u^{-1/3} \,  \frac{d^{2} \Psi}{df^{2}}\,. 
\ee
Since $\Psi = \Psi_{\GR} + \beta \eta^{d} u^{b}$, we find that 
\be
\dot{E}  = \dot{E}_{\GR} \left[ 1 + \pi^2 {\cal{M}}^2 \beta \eta^d b \left(b-1\right) u^{b-2} \left(\frac{d^{2} \Psi_{\GR}}{df^{2}}\right)^{-1} \right]\,,
\label{dotE-phase-ppE2}
\ee
where $\dot{E}_{\GR}$ can be written in terms of the GW phase as 
\be
\dot{E}_{\GR}  = - \frac{1}{6} \, \dot{f}_{\GR}^{2} \,  {\cal{M}}^{2} \,  u^{-1/3} \, \frac{d^{2} \Psi_{\GR}}{df^{2}}\,.
\ee

This shows that measurement of $\dot{E}$ from a {\it circular} binary
pulsar would allow us to constrain both the amplitude and the phase
ppE parameters. All known binary pulsars, however, are in eccentric
orbits. The $\dot{f}_{\GR}$ used in Eqs.~\eqref{dotE-amp-ppE}
and~\eqref{dotE-phase-ppE}, or equivalently the $\dot{E}_{\GR}$ used
in Eqs.~\eqref{dotE-amp-ppE2} and~\eqref{dotE-phase-ppE2}, must be
that of an eccentric orbit, namely~\cite{1963PhRv..131..435P}
\be
\dot{E}_{\GR} = - \frac{32}{5} \eta^2  \frac{m^5}{a^{5}} \left(1 - e^{2} \right)^{-7/2} \left(1 + \frac{73}{24} e^{2} + \frac{37}{96} e^{4} \right)\,. 
\ee
The ppE corrections to $\dot{E}$ computed in
Eqs.~\eqref{dotE-amp-ppE2} and~\eqref{dotE-phase-ppE2} are built from
circular, non-spinning ppE templates, the only ones currently
available. The ppE corrections to the GW luminosity computed here are
really only the leading order terms in a post-circular
expansion~\cite{Yunes:2009yz}, i.e.,~an expansion in small
eccentricity $e$. If the eccentricity is small, such as with PSR
J0737-3039 ($e = 0.088$), such an approximation is well-justified and
will not strongly affect the constraints we place on the ppE
framework.

{\emph{Orbital Period Decay}}. The GW luminosity enters into binary
pulsar observables through the orbital decay: ${\dot{P}}/{P} =
({3}/{2}) \, {\dot{E}_{b}}/{E_{b}} = -({3}/{2}) \, {\dot{E}}/{E_{b}}$,
where in the second equality we used energy balance: the amount of
binding energy lost by the system is equal to minus the amount of
energy carried away by GWs, $\dot{E}_{b} = - \dot{E}$. Using
Eq.~\eqref{dotE-amp-ppE2} and~\eqref{dotE-phase-ppE2}, we then find
that the $\dot{P}$ corrected by amplitude ppE parameters is
\be
\frac{\dot{P}}{P} = \left(\frac{\dot{P}}{P}\right)_{\GR} \left(1 + 2 \, \alpha \,  \eta^{c} u^{a} \right)\,,
\label{Pdot-amp}
\ee
while that corrected by phase ppE parameters is 
\be
\label{Pdot-phase}
\frac{\dot{P}}{P} = \left(\frac{\dot{P}}{P}\right)_{\GR} \left(1 + \frac{48}{5} \, \beta \, \eta^{d} \, b (b -1) \, u^{b+5/3}\right)\,.
\ee
The quantity $(\dot{P}/{P})_{\GR}$ stands for the orbital decay in GR
for an eccentric inspiral, namely~\cite{1963PhRv..131..435P}
\be
\left(\frac{\dot{P}}{P}\right)_{\GR} =  -\frac{96}{5} \frac{\eta \, m^3}{a^{4}} \left(1 - e^{2} \right)^{-7/2} \left(1 + \frac{73}{24} e^{2} + \frac{37}{96} e^{4} \right)\,.
\ee
Recall again that the ppE corrections [the second terms inside the
  parenthesis of Eqs.~\eqref{Pdot-amp} and~\eqref{Pdot-phase}], are
only valid to leading order in the post-circular approximation.  In
deriving these expressions, we have used the fact that the observed
$\dot{P}/P$ is very close to the GR value: $({\dot{P}}/{P})_{\rm obs}
= ({\dot{P}}/{P})_{\GR} (1 + \delta )$.  The observational error
$\delta \equiv (\delta \dot{P})/\dot{P} \ll 1$, meaning that the error
on $\dot{P}$ dominates over the error on $P$.

Since binary pulsar observations have confirmed GR up to observational
error, we can now place relation constraints on the ppE
framework. Focusing first on the amplitude ppE parameters, we find
that
\be
\left|\alpha\right| \leq \frac{1}{2} \frac{\delta}{\eta^{c} u^{a}} \,.
\label{alpha-prior}
\ee
For the phase ppE parameter,
\be
\left|\beta\right| \leq \frac{5}{48 \, |b| \, \left|b - 1\right|} \frac{\delta}{\eta^{d} u^{b+5/3}}\,.
\label{beta-prior}
\ee
A binary pulsar measurement of $\dot{P}$ to an accuracy $\delta$
allows us to constrain $\alpha$ and $\beta$, given some value for
$(a,b,c,d)$, the symmetric mass ratio and the GW frequency, or
equivalently, the orbital period.

Before proceeding, let us first discuss the apparent degeneracy
between the amplitude and the phase correction. Comparing
Eqs.~\eqref{Pdot-amp} and~\eqref{Pdot-phase}, one realizes that if
changes to the GW amplitude and phase are due to the {\emph{same}}
mechanism (for example, if only $\dot{E}$ is modified), then we must
have $a = b + 5/3$, $c = d$, and $\beta = 5 \alpha/[48 b (b -
  1)]$. The ppE scheme, however, allows for modifications to
{\emph{both}} the dissipative ($\dot{E}$) sector and the conservative
($E_{b}$) sector.  If both sectors are modified, there will be two
sets of independent modifications, one to the phase and one to the
amplitude. If a ppE correction is introduced to the GW amplitude, then
it is constrained by Eq.~\eqref{Pdot-amp}; if a ppE correction is
introduced to the GW phase, then it is constrained by
Eq.~\eqref{Pdot-phase}. These constraints on the amplitude and phase
ppE parameters are thus {\emph{independent}} from each other, even
though a constraint or measurement of them would not allow a
one-to-one mapping to a conservative or dissipative
modification. Thus, conservative and dissipative modifications are in
fact degenerate, even though the phase and amplitude measurements are
not.

{\emph{Binary Pulsar Constraint}}. Let us now employ the recent
measurements of~\cite{Lyne:2004cj,Kramer:2006nb} on PSR J0737-3039 to
constrain $(\alpha,\beta)$. This binary consists of two neutron stars
with component masses $m_{1} = 1.3381(7) \, M_{\odot}$ and $m_{2} =
1.2489(7) \, M_{\odot}$ in an almost circular orbit with eccentricity
$e = 0.0877775(9)$ and period $P = 8834.535000(4) \; {\rm{s}}$. The
symmetric mass ratio is $\eta \simeq 0.24970$, the chirp mass is
${\cal{M}} \simeq 5.5399 \times 10^{-6} \; {\rm{s}}$, the GW frequency
is $f = 2/P \simeq 2.263842976 \times 10^{-4} \; {\rm{Hz}}$, and the
reduced frequency is $u \simeq 3.940046595 \times 10^{-9}$. The time
derivative of the period is measured to be $\dot{P} = -1.252(17)
\times 10^{-12}$, which implies an uncertainty of $\delta = 0.017
\times 10^{-12}/(1.252 \times 10^{-12}) \simeq 10^{-2}$. This
uncertainty is comparable to the systematic error in the ppE
parameters due to the neglect of eccentricity effects in PSR
J0737-3039, which roughly scale as $e^{2} \simeq 0.0077$. An increase
in the accuracy of the $\dot{P}$ measurement, reducing $\delta$, would
not allow us to place stronger constraints until the ppE templates are
extended to include eccentricity.

\begin{figure*}[ht]
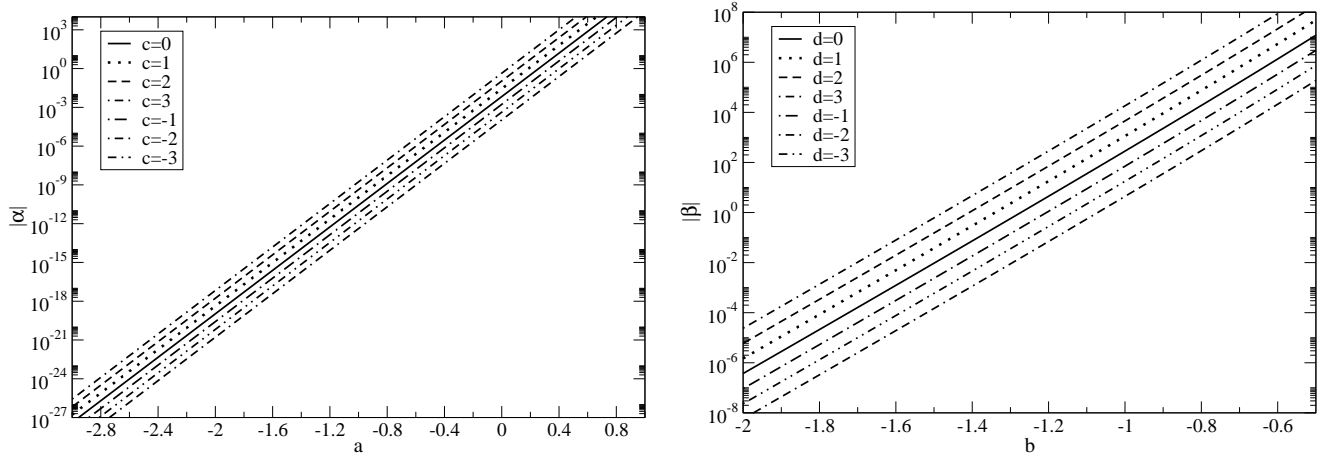

\begin{center}
\begin{tabular}{cc}
\epsfig{file=alpha-a.eps,width=8.5cm,clip=true} \quad
\epsfig{file=beta-b.eps,width=8.5cm,clip=true}
\end{tabular}
\end{center}
\caption{\label{alpha-beta-cons} Left: Constraint on $|\alpha|$ as a function of $a$ for fixed $c$. Right: Constraint on $|\beta|$ as a function of $b$ for fixed $d$. The area below the curves is allowed, while the area above is ruled out.}
\end{figure*}

Figure~\ref{alpha-beta-cons} plots the double binary pulsar
constraints on $(|\alpha|,|\beta|)$ as a function of the exponent ppE
parameters $(a,b)$ for fixed $(c,d)$. The area above the curves is
excluded by binary pulsar observations, forcing $(\alpha,\beta)$ to be
smaller than a value which depends on $(a,b,c,d)$.  Generally, if $a <
-0.4$ then $|\alpha| < 10^{-6}$ for all plotted values of $c$, while
if $b<-1.9$ then $|\beta| < 10^{-6}$ for all plotted values of
$d$. For $a > 0.2$ and $b> -4/3$, $\alpha$ and $\beta$ can be greater
than unity for all plotted values of $(c,d)$. This makes sense: as
$(a,b)$ become large and positive, the ppE correction becomes smaller
for low reduced frequency sources.

These constraints are consistent with other constraints on GR
deviations from binary pulsars. For example, one can place a generic
constraint on the time-variation of Newton's constant $G$ with a
binary pulsar
observation~\cite{1988PhRvL..61.1151D,1994ApJ...428..713K}: $\dot{G}/G
\leq (\delta P)/(2P)$, where $\delta P$ is whatever part of $\dot{P}$
that is otherwise unexplained. Using PSR
J0737-3039~\cite{Lyne:2004cj,Kramer:2006nb} one infers that $\dot{G}/G
< 3 \times 10^{-11} \; {\rm{yr}}^{-1}$. Allowing for Newton's constant
to be a linear function of time leads to a modification that can be
mapped to Eq.~\eqref{ppE-h} with $|\alpha| = (5/512) (\dot{G}/G) M$,
$c = 3/5$ and $a = -8/3$ for the amplitude parameters, and
$|\beta|=(25/65536) (\dot{G}/G) M$, $d=3/5$ and $b=-13/3$ for the
phase parameters~\cite{Yunes:2009bv}. From the binary pulsar
constraint on $\dot{G}/G$, we then infer that $|\alpha| \lesssim
10^{-25}$ and $|\beta| \lesssim 10^{-27}$, which is consistent with
Eqs.~\eqref{alpha-prior} and~\eqref{beta-prior} and
Fig.~\ref{alpha-beta-cons}.

Our constraints on $\alpha$ look extremely strong (e.g.,~for $a < -2$,
then $|\alpha| \lesssim 10^{-20}$).  However, this does not imply that
the unconstrained region (below the curves in
Fig.~\ref{alpha-beta-cons}) is uninteresting. For example,
constraining $\dot{G}/G$ below $10^{-12} \; {\rm{yr}}^{-1}$ or
$10^{-13} \; {\rm{yr}}^{-1}$ implies constraining $|\alpha|$ below
$10^{-25}$. This is interesting as there are GR modifications that
suggest $\dot{G}/G$ deviations of this order may be
present~\cite{Melnikov:2009zza}. On the other hand, the smallness of
the y-axis of Fig.~\ref{alpha-beta-cons} does suggest that $\alpha$
and $\beta$ are perhaps not the best ``coordinates'' with which to
measure GR deviations when $a$ and $b$ are sufficiently negative.

We conclude this discussion with some caveats on the constraints we
find. First, we have here neglected the effect of eccentricity in the
ppE correction, although this is accounted for in the GR part of
$\dot{P}/P$. We have not studied eccentric ppE templates because these
do not yet exist.  This is due to the difficulty in constructing
analytically simple Fourier transforms of eccentric inspiral waveforms
in the stationary phase approximation~\cite{Yunes:2009yz}. Second,
because of the ppE degeneracy between conservative and dissipative
corrections, we have only examined dissipative ones here; recall,
however, that amplitude and phase measurements are truly
independent. Since the conservative sector can be thought of as
unmodified, this allows us to use the GR measured values for the
components' masses, as the Shapiro time delay and periapsis precession
are the same as in GR (leading to identical results for $m_1$ and
$m_2$ in GR and in the ppE extended theory). We could have instead
allowed for both conservative and dissipative corrections, and then
analyzed how these affect {\emph{all}} binary pulsar observables. A
combined analysis of all these effects will presumably lead to a
stronger bound on the ppE parameters; we leave this to future work. We
note that we could have studied constraints on ppE from Solar System
observations. However, the exquisite accuracy of the double binary
pulsar measurements, and the fact that this is a much stronger-field
source than any Solar System one, means that Solar System constraints
will not be as stringent as the ones discussed here.

{\emph{Implications for GW Data Analysis}}. Once GWs are detected, one
would like to implement the ppE framework in a realistic data analysis
pipeline. Such a pipeline will likely employ techniques from Bayesian
analysis~\cite{Cornish:2005qw}, which relies heavily on the priors
chosen for the parameters searched over. The prior tells us whether
certain regions of parameter space are allowed or likely to occur in
Nature. The priors for the ppE parameters should be constructed
following current Solar System and binary pulsar
constraints. Equations~\eqref{alpha-prior} and~\eqref{beta-prior}
represent the most stringent prior found to date for these parameters
using binary pulsar observations.

{\emph{Acknowledgments}}.~We are grateful to Frans Pretorius, Neil Cornish and the GW group 
at the University of Wisconsin, Milwaukee for hosting the ``GW Tests of Alternative
Theories of Gravity in the Advanced Detector Era'' workshop, where
this paper was conceived. NY acknowledges support from NSF grant
PHY-0745779; SAH acknowledges support from NSF Grant PHY-0449884.

\bibliography{review}

\end{document}